\documentclass[journal=jctcce,manuscript=article,layout=twocolumn]{achemso}
\setkeys{acs}{doi=true}

\usepackage[version=3]{mhchem}
\usepackage{booktabs}
\usepackage{braket}
\usepackage{bm}
\usepackage{hyperref}
\usepackage{cleveref}
\usepackage{subcaption}

\mciteErrorOnUnknownfalse

\newcommand*{\pyscf}{\textsc{PySCF}}
\newcommand*{\pyflosic}{\textsc{PyFLOSIC2}}
\newcommand*{\pyflosicgui}{\textsc{PyFLOSIC2:GUI}}
\newcommand*{\libxc}{\textsc{Libxc}}
\newcommand*{\fodmc}{\textsc{fodMC}}
\newcommand*{\chimerax}{\textsc{ChimeraX}}
\newcommand*{\plotly}{\textsc{Plotly}}
\newcommand*{\matplotlib}{\textsc{Matplotlib}}
\newcommand*{\eminus}{\textsc{eminus}}
\newcommand*{\chillipy}{\textsc{chilli.py}}
\newcommand*{\python}{\textsc{Python}}

\newcommand*{\gitlab}{\textsc{GitLab}}
\newcommand*{\psifour}{\textsc{Psi4}}
\newcommand*{\citeref}[1]{ref.~\citenum{#1}}
\newcommand*{\citerefs}[1]{refs.~\citenum{#1}}

\usepackage{xcolor}

\newcommand*\citeit[1]{\citeauthor{#1}\cite{#1}}

\author{Sebastian Schwalbe}
\email{s.schwalbe@hzdr.de}
\affiliation{Contributed equally to this work}
\alsoaffiliation{Center for Advanced Systems Understanding (CASUS), D-02826 G{\"o}rlitz, Germany}
\alsoaffiliation{Helmholtz-Zentrum Dresden-Rossendorf (HZDR), D-01328 Dresden, Germany}

\author{Wanja Timm Schulze}
\affiliation{Contributed equally to this work}
\alsoaffiliation{Institute for Physical Chemistry, Friedrich Schiller University, D-07743 Jena, Germany}

\author{Kai Trepte}
\affiliation{Taiwan Semiconductor Manufacturing Company North America, San Jose, USA}

\author{Susi Lehtola}
\email{susi.lehtola@alumni.helsinki.fi}
\affiliation{Department of Chemistry, University of Helsinki, P.O. Box 55, FI-00014 Helsinki, Finland}

\title{Ensemble Generalization of the Perdew--Zunger Self-Interaction Correction: a Way Out of Multiple Minima and Symmetry Breaking}

\keywords{SIC, multiple minima, symmetry breaking}

\begin{document}

\begin{tocentry}
\includegraphics[height=1.75in]{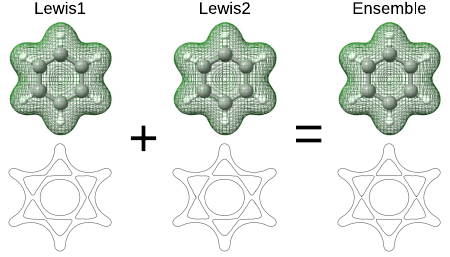}
\end{tocentry}

\begin{abstract}

The Perdew--Zunger (PZ) self-interaction correction (SIC) is an established tool to correct unphysical behavior in density functional approximations.
Yet, PZ-SIC is well-known to sometimes break molecular symmetries.
An example of this is the benzene molecule, for which PZ-SIC predicts a symmetry-broken electron density and molecular geometry, since the method does not describe the two possible Kekul{\'e} structures on an even footing, leading to local minima [Lehtola et al, J. Chem. Theory Comput. 2016, 12, 3195].
PZ-SIC is often implemented with Fermi--L{\"o}wdin orbitals (FLOs), yielding the FLO-SIC method, which likewise has issues with symmetry breaking and local minima [Trepte et al, J. Chem. Phys. 2021, 155, 224109].

In this work, we propose a generalization of PZ-SIC---the ensemble PZ-SIC (E-PZ-SIC) method---which shares the asymptotic computational scaling of PZ-SIC (albeit with an additional prefactor).
E-PZ-SIC is straightforwardly applicable to various molecules, merely requiring one to average the self-interaction correction over all possible Kekul{\'e} structures, in line with chemical intuition.
We showcase the implementation of E-PZ-SIC with FLOs, as the resulting E-FLO-SIC method is easy to realize on top of an existing implementation of FLO-SIC.
We show that E-FLO-SIC indeed eliminates symmetry breaking, reproducing a symmetric electron density and molecular geometry for benzene. 
The ensemble approach suggested herein could also be employed within approximate or locally scaled variants of PZ-SIC and their FLO-SIC versions.
\end{abstract}

\section{Introduction \label{sec:intro}}

Given their high predictive power and easy applicability combined with a reasonable level of computational effort, calculations within density functional theory~\cite{Hohenberg1964_PR_864, Kohn1965_PR_1133} (DFT) are nowadays extensively leveraged for a wide range of studies across various fields of computational chemistry, physics, and materials science,\cite{Orio2009_PR_443, Neugebauer2013_WIRCMS_438, Mardirossian2017_MP_2315}
as is illustrated by some of our recent studies in chemical science\cite{Schulze2023}  and warm dense matter.\cite{Moldabekov2023, Moldabekov2024, Dornheim2024}

In addition to their direct use to understand the behavior of matter in various situations, DFT calculations have also had a pivotal role in enabling the use of machine learning (ML) methods for material modeling by allowing the generation of large sets of reference data with a consistent level of accuracy.\cite{Schleder2019_JPM_32001, Huang2023_S_170}
Such data can be used to train ML models that enable the simulation of complex systems outside the reach of direct DFT calculations at an accuracy comparable to that of the DFT calculations used to train the ML model.\cite{Fiedler2022_PRM_40301, Huang2023_S_170}

Although DFT is exact in theory, practical implementations of DFT rely on so-called density functional approximations (DFAs), since the exact exchange-correlation functional that describes the quantum mechanical many-body interactions of electrons with each other remains unknown.
Unfortunately, a variety of issues exist in presently available DFAs; any issues with a given DFA complicate computational modeling and will also be inherited by ML models trained to data computed with that DFA.

For example, recent work has shown that the numerical well-behavedness\cite{Lehtola2022_JCP_174114, Lehtola2023_JPCA_4180, Lehtola2023_JCTC_2502} (or even the reproducibility!\cite{Lehtola2023_JCP_114116}) of DFAs has not been traditionally given adequate consideration by their developer community, even though such issues can already be observed in atomic calculations: many recent DFAs do not allow one to converge the total energy for a single atom to the complete basis set limit even with fully numerical\cite{Lehtola2019_IJQC_25968} methods.\cite{Lehtola2022_JCP_174114, Lehtola2023_JPCA_4180, Lehtola2023_JCTC_2502, Lehtola2023_JCP_114116, Bjoergve2024_JCP_162502}

Naturally, numerical ill-behavior in DFAs also leads to artifacts in observables: for instance, the spurious oscillations along nuclear displacements found by \citeauthor{Sitkiewicz2022_JPCL_5963} in \citerefs{Sitkiewicz2022_JPCL_5963} and \citenum{Sitkiewicz2024_JCTC_3144} that lead to significant errors in harmonic and anharmonic infrared and Raman frequencies and intensities likely arise from numerical ill-behavior of the DFA that may be found already in the above-discussed types of atomic calculations.
Note that any properly fit ML model would likewise reproduce such spurious oscillations; the quality of any ML model is limited by the data used to train it.

The focus of this work is another shortcoming of  DFAs, which also affects several recent DFAs: self-interaction error (SIE)---the artificial interaction of electrons with themselves---has been known for a long time and is still considered a major issue in practical DFT calculations.\cite{Bao2018_JPCL_2353} 
SIE leads to qualitatively incorrect wave functions in many kinds of systems, such as artificially unbound anions like \ce{F-}\cite{Galbraith1996_JCP_862} and too delocalized defect states \cite{Pacchioni2008_JCP_182505}.

The standard way to address the SIE in practical DFT calculations is to employ hybrid functionals, which are admixtures of semi-local DFAs with exact exchange.
Various hybrid functionals have been developed, ranging from global mixtures\cite{Becke1993_JCP_5648} to range-separated\cite{Gill1996_MP_1005, Leininger1997_CPL_151} and local mixtures\cite{Jaramillo2003_JCP_1068} of the exact exchange energy.
While including exact exchange decreases SIEs, since no successful functional employs 100\% exact exchange, \cite{Mardirossian2017_MP_2315} hybrid functionals are not impervious to the SIE which thereby continues to be an issue in DFT calculations.
In addition, the inclusion of exact exchange is well-known to be problematic for metallic systems, for example, motivating investigations of alternative approaches.

\citeit{Perdew1981_PRB_5048} (PZ) proposed already back in \citeyear{Perdew1981_PRB_5048} a self-interaction correction (SIC) that removes the SIE estimated by the sum of DFA errors of one-electron densities that sum up to the total electron density; the SIE of one-electron systems with modern DFAs has been recently assessed in \citerefs{Schwalbe2022_JCP_174113} and \citenum{Lonsdale2023_JCP_44102}.
Although the idea of PZ-SIC is elegant, it does not always lead to improvement over the base DFA and can even lead to worse agreement with experiment.\cite{Jonsson2015_PCS_1858, Perdew2015__193}
However, it has also been used with some resounding success: for instance, PZ-SIC correctly describes Al-dopant in $\alpha$ quartz\cite{Gudmundsdottir2015_NJP_83006} and molecular Rydberg states,\cite{Sigurdarson2023_JCP_214109} while uncorrected DFAs fail to achieve reliable accuracy in these applications.

PZ-SIC calculations are considerably more complicated to carry out than DFT calculations, because the orbital-by-orbital SIC breaks the usual invariances\cite{Lehtola2020_M_1218} observed in DFT, the functional now depending explicitly on the individual occupied orbitals instead of only the electron density as in the case of the Kohn--Sham functional.\cite{Pederson1984_JCP_1972, Lehtola2014_JCTC_5324}
This complication has led to many flavors and approximations of PZ-SIC.

To start, the original approach of PZ-SIC by \citeit{Pederson1984_JCP_1972} was based on real-valued orbitals (RSIC)\cite{Pederson1984_JCP_1972, Pederson1985_JCP_2688, Pederson1986_PHD}.
The complex-valued orbital SIC
(CSIC) first studied by \citeit{Kluepfel2011_PRA_50501} leads to lower total energies than those obtained with real-valued orbitals;\cite{Kluepfel2011_PRA_50501, Kluepfel2012_JCP_124102, Lehtola2014_JCTC_5324, Lehtola2015_JCTC_5052, Lehtola2015_JCTC_839}
in fact, it has been shown that complex-valued orbitals are necessary to minimize the PZ-SIC functional, as real-valued orbitals are merely saddle points on the complex orbital manifold.\cite{Lehtola2016_JCTC_3195}
Perdew and coworkers attributed the importance of the complex orbitals to the role of noded electron densities in PZ-SIC.\cite{Shahi2019_JCP_174102}

RSIC and CSIC involve the optimization of a $N \times N$ unitary matrix\cite{Lehtola2013_JCTC_5365} for $N$ occupied orbitals for an objective function that is similar to the one used in the Edmiston--Ruedenberg\cite{Edmiston1963_RMP_457} orbital localization method.
It is then not so surprising that the optimal RSIC and CSIC orbitals turn out to be localized, as well.\cite{Pederson1985_JCP_2688}

A variety of orbital localization methods based on unitary optimization has been suggested in the literature for various purposes.
In addition to the Edmiston--Ruedenberg\cite{Edmiston1963_RMP_457} method already mentioned, others include Foster--Boys localization,\cite{Foster1960_RMP_300} the von Niessen method,\cite{Niessen1972_JCP_4290} fourth moment localization,\cite{Hoyvik2012_JCP_224114}  and Pipek--Mezey localization\cite{Pipek1989_JCP_4916} as well as its generalized variant\cite{Lehtola2014_JCTC_642} that has recently also been extended to the generation of maximally localized Wannier functions.\cite{Jonsson2017_JCTC_460, Clement2021_JCTC_7406, Schreder2024_JCP_214117}
While these methods do not variationally minimize the PZ-SIC functional, any localized orbitals do yield better estimates for the value of the PZ-SIC functional than that obtained with the delocalized canonical orbitals.
We note that such an approximate implementation of PZ-SIC based on localized orbitals from the selected columns of the density matrix (SCDM) orbital localization method\cite{Damle2015_JCTC_1463, Fuemmeler2023_JCTC_8572} has been recently described by \citeit{Peralta2024}.

On a different path, Luken and coworkers discovered a way to generate localized orbitals based on the Fermi hole.\cite{Luken1982_TCA_265, Luken1984_TCA_279}
Given a point in three-dimensional space called a Fermi-orbital descriptor (FOD), all the electron density at that point can be associated with a localized orbital,\cite{Luken1982_TCA_265} which is commonly called a Fermi orbital.
Choosing $N$ FODs for $N$ orbitals, one then obtains a set of localized orbitals; in \citeyear{Luken1984_TCA_279} \citeit{Luken1984_TCA_279} suggested orthogonalizing them with the method of \citeit{Loewdin1950_JCP_365} to obtain a set of orthonormal localized orbitals, usually referred to as Fermi--L{\"o}wdin orbitals (FLOs).

These developments led to the suggestion of FLO-SIC in \citeyear{Pederson2014_JCP_121103}: the orbital localization based on unitary optimization in RSIC or CSIC is replaced by the use of FLOs in FLO-SIC.\cite{Pederson2014_JCP_121103, Pederson2015_JCP_64112, Pederson2015__153, Yang2017_PRA_52505}
A considerable reduction in the number of optimizable parameters is achieved in FLO-SIC.
While CSIC and RSIC require the optimization of $N(N-1)$ and $N(N-1)/2$ orbital rotation angles, respectively, in FLO-SIC only $3N$ parameters need to be optimized.
Although FLO-SIC therefore has considerably fewer parameters for $N\gg 1$, computing the FOD derivatives is more expensive than computing the derivatives of the unitary rotation angles in CSIC or RSIC,\cite{Schwalbe2020_JCP_84104} and this step is one of the typical bottlenecks of FLO-SIC calculations.

Because of the explicit dependence on the representation of the individual occupied orbitals with significant localized character, both RSIC and CSIC have been shown to exhibit multiple electronic local minima, which often correspond to different Kekul{\'e} structures.\cite{Lehtola2016_JCTC_3195}
Unsurprisingly, the problem of PZ-SIC with local minima also persists in FLO-SIC, as we have recently shown in \citeref{Trepte2021_JCP_224109}; the tendency of FLO-SIC to converge to various local minima depending on the used initial guess has also been discussed by others for diverse systems.\cite{Hahn2017_JCTC_5823, Kao2017_JCP_164107, Pederson2023_JCP_84101, Hooshmand2023_JCP_234121, Maniar2024_JCP_144301}

This problem with local minima is a significant complication for practical calculations with PZ-SIC and FLO-SIC, because the solutions corresponding to the distinct local minima often exhibit severe symmetry breaking, leading to qualitatively incorrect geometries and/or large artifactual dipole moments.\cite{Lehtola2016_JCTC_3195, Trepte2021_JCP_224109}
In this work, we suggest that the problem of local minima can be removed in PZ-SIC and FLO-SIC with an ensemble approach by averaging the self-interaction correction over various sets of localized orbitals, yielding the ensemble PZ-SIC (E-PZ-SIC) and E-FLO-SIC methods, respectively.

Showcasing the E-FLO-SIC method with the benzene (\ce{C6H6}) molecule that is known to be problematic for CSIC,\cite{Lehtola2016_JCTC_3195} RSIC,\cite{Lehtola2016_JCTC_3195} as well as FLO-SIC\cite{Hahn2017_JCTC_5823, Trepte2021_JCP_224109}, we demonstrate that while FLO-SIC exhibits symmetry breaking of either the electron density or of the spin density for calculations employing FOD geometries following Lewis~\cite{Lewis1916_JACS_762} and Linnett~\cite{Linnett1960_N_859, Linnett1961_JACS_2643, Luder1966_JCE_55} theory, respectively,\cite{Trepte2021_JCP_224109} the E-FLO-SIC method of this work correctly reproduces a symmetric electron density and zero spin density using an ensemble of either type of electronic geometries.

To continue, the symmetry breaking of the electron density in FLO-SIC following a Lewis structure has been shown to lead to non-negligible bond length alternation in the molecule,\cite{Trepte2021_JCP_224109} in line with earlier results for RSIC and CSIC.\cite{Lehtola2016_JCTC_3195}
In other words, these methods do not predict benzene to be aromatic, while FLO-SIC using a Linnett structure results in broken spin symmetry as already mentioned.\cite{Trepte2021_JCP_224109}.
In contrast, we show that the E-FLO-SIC method predicts a symmetric ground state geometry for benzene without breaking spin symmetry. 
We also show that the proposed E-FLO-SIC method reproduces the same results whether Lewis or Linnett structures are used in the calculations and that various starting points for the optimization lead to convergence to the same minimum.

The layout of this work is as follows.
Next, in \cref{sec:theory}, we describe the theory for the proposed ensemble approach.
Computational details on the implementation are given in \cref{sec:compdec}.
The results of the application of the method onto the benzene molecule are discussed in \cref{sec:results}.
The article concludes with a brief summary and discussion in \cref{sec:summary}.
Atomic units are used throughout the manuscript unless specified otherwise.

\section{Theory \label{sec:theory}}

The Kohn--Sham~\cite{Kohn1965_PR_1133} (KS) total energy $E^{\text{KS}}$ is given by
\begin{align}
    E^{\text{KS}}[n_{\alpha},n_{\beta}] &= T_{\text{s}}[n_{\alpha},n_{\beta}] + V_\text{ext}[n] \nonumber \ \\ 
                                         &\phantom{=\,} + E_{\text{J}}[n] + E_{\text{XC}}[n_{\alpha},n_{\beta}],
                                         \label{eq:Eks}
\end{align}
where $T_{\text{s}}$ is the kinetic energy of the non-interacting system, $V_\text{ext}$ is the external potential energy, $E_{\text{J}}$ is the Coulomb energy, $E_{\text{XC}}$ is the exchange-correlation energy of the employed DFA, $n$ is the total electron density, and $n_{\alpha}$ and $n_{\beta}$ are the electron densities for spin $\alpha$ and $\beta$, respectively.
The PZ formalism rectifies the KS total energy estimate arising from a DFA by removing the SIE from \cref{eq:Eks}
\begin{align}
  E^{\text{PZ}}[n_{\alpha},n_{\beta}] &= E^{\text{KS}}[n_\alpha,n_\beta] - E_{\text{SIE}}^{\text{PZ}}[n_{\alpha},n_{\beta}].\label{eq:PZ-tot}
\end{align}
The SIE is estimated in the PZ approach by the orbital-by-orbital ansatz
\begin{align}
      E^{\text{PZ}}_\text{SIE}[n_\alpha, n_\beta] =& \sum_{\sigma} \sum_{i=1}^{N_{\rm el}^{\sigma}} \Big(E_\text{J}[n_{i,\sigma},0] \nonumber \ \\
      & + E_\text{XC}[n_{i,\sigma},0]\Big) \label{eq:pz-corr},
\end{align}
where $N_{\rm el}^{\sigma}$ is the number of electrons in spin channel $\sigma$, and $n_{i,\sigma}$ denotes the density arising from the $i$th spin $\sigma$ orbital, which sum to the total spin $\sigma$ density, $\sum_{i=1}^{N_{\rm el}^\sigma} n_{i,\sigma} = n_\sigma$.

The issues with local minima in PZ-SIC and FLO-SIC have been identified by \citeit{Lehtola2016_JCTC_3195} and \citeit{Trepte2021_JCP_224109} to arise in situations where more than one set of localized orbitals (FODs in FLO-SIC) yields a local minimum of the electronic energy.
The issues arise because the correction in \cref{eq:pz-corr} is made more negative when the orbital density is localized, which leads to the system adopting a single Kekul{\'e} structure, thus resulting in spontaneous symmetry breaking.
Relaxing the electron density then enhances the bias towards this single Kekul{\'e} structure, increasing the amount of symmetry breaking.
This symmetry breaking can result in artifactual dipole moments as well as qualitative errors in molecular geometries, such as bond length alternation in benzene.\cite{Lehtola2016_JCTC_3195, Trepte2021_JCP_224109}

To address this problem, we propose using an ensemble of electronic configurations to evaluate the SI correction, instead.
In our proposed approach, instead of \cref{eq:pz-corr}, the self-interaction correction is determined using an ensemble of sets of localized orbitals, i.e. different Kekul{\'e} structures
\begin{align}
      E^{\text{Ensemble}}_\text{SIE}[n_\alpha, n_\beta] =& \sum_{\sigma} \frac{1}{N_{\rm conf}^\sigma}  \sum_{P=1}^{N_{\rm conf}^{\sigma}} \sum_{i}^{N_{\rm el}^{\sigma}} \Big(E_\text{J}[n_{i,\sigma}^{P},0] \ \nonumber \\ 
      &+ E_\text{XC}[n_{i,\sigma}^{P},0]\Big),  \label{eq:ensemble-si}
\end{align}
where $N_{\rm conf}^\sigma$ is the number of configurations in the ensemble for spin $\sigma$.
The total energy is then estimated in analogy to \cref{eq:PZ-tot} by employing the ensemble-averaged SIE  defined by \cref{eq:ensemble-si} to correct the KS total energy of \cref{eq:Eks}
\begin{align}
  E^{\text{Ensemble}}[n_{\alpha},n_{\beta}] &= E^{\text{KS}}[n_\alpha,n_\beta] \nonumber\ \\
                                            &\phantom{=\,} - E_{\text{SIE}}^{\text{Ensemble}}[n_{\alpha},n_{\beta}]. \label{eq:EFLOSIC-tot}
\end{align}
We already observe from these results that E-PZ-SIC maintains the same asymptotic scaling as PZ-SIC with an additional prefactor of $N_{\rm conf}^\sigma$ for the number of necessary Kekul{\'e} structures.

Although the proposed method can also be applied in the context of RSIC and CSIC, it is remarkably simple to implement in the context of FLO-SIC, because the optimization of the electron density and the local orbitals are decoupled in FLO-SIC;\cite{Schwalbe2020_JCP_84104} we therefore demonstrate the method with E-FLO-SIC calculations.
This has the added benefit that the identification of the various Kekul{\'{e}} structures is simple, because the localization of the electrons in the various structures can be observed from the FODs.\cite{Schwalbe2019_JCC_2843}

We will show in \cref{sec:results} that when the various configurations in the ensemble try to break the symmetry of the electron density in complementary directions, the net effect is zero symmetry breaking, while the major effects of the self-interaction correction are preserved.

Because the ansatz in \cref{eq:ensemble-si} is linear in the ensemble, the modifications to the self-consistent field (SCF) equations used to determine the optimal electron density in E-FLO-SIC are trivially obtained from known results for FLO-SIC.
While in the FLO-SIC method the effective Fock matrix is given by\cite{Yang2017_PRA_52505, Schwalbe2020_JCP_84104}
\begin{align}
    \boldsymbol{F}_{\text{FLO-SIC}}^{\sigma} &= \boldsymbol{F}_{\text{KS}}^{\sigma} - \boldsymbol{F}_{\text{SIC}}^{\sigma}, \label{eq:f-flosic}
\end{align}
where $\boldsymbol{F}_{\text{KS}}^{\sigma}$ is the Kohn--Sham Fock matrix and $\boldsymbol{F}_{\text{SIC}}^{\sigma}$ is its correction, in the ensemble method of this work the correction to the Fock matrix is simply given by the ensemble average of the corrections
\begin{equation} \label{eq:f-eflosic}
\begin{split} 
    \boldsymbol{F}_{\text{Ensemble}}^{\sigma} &= \boldsymbol{F}_{\text{KS}}^{\sigma} -  \frac{1}{N_{\rm conf}^\sigma} \left( \sum_{P=1}^{N_{\rm conf}^\sigma} \boldsymbol{F}_{\text{SIC}}^{P,\sigma}\right) \\  
    &= \boldsymbol{F}_{\text{KS}}^{\sigma} - \boldsymbol{F}_{\text{SIC}}^{\text{Ensemble},\sigma}. 
\end{split}
\end{equation}

As can be seen from the above, the proposed E-FLO-SIC method works analogously to conventional FLO-SIC; the only difference is that an ensemble-averaged self-interaction correction is employed in E-FLO-SIC.
Importantly, the FOD optimizations for each member of the ensemble are independent, and can therefore be carried out in parallel for a fixed electron density.
Once the FODs of all configurations have been optimized, the ensemble SIE $E_{\text{SIE}}^{\text{Ensemble}}$ and the ensemble SIC Hamiltonian $\boldsymbol{F}_{\text{SIC}}^{\text{Ensemble},\sigma}$ can easily be computed, and the electron density relaxed with the SCF method.

The analogous RSIC or CSIC implementation of E-PZ-SIC would be based on a set of common occupied orbitals to represent the total density, and a set of unitary matrices representing the localized orbitals in each Kekul{\'e} structure.
As in E-FLO-SIC, the unitary matrix optimizations in E-PZ-SIC are independent and can be performed in parallel, while the optimization of the total density would include a correction averaged over all Kekul{\'e} structures.

\section{Computational Details \label{sec:compdec}}

For the sake of transparency and reproducibility of the results, this manuscript follows the free and open-source software (FOSS) approach~\cite{Lehtola2022_WIRCMS_1610}, and exclusively employs open-source codes to enable verification of the implemented model and results.
We implemented E-FLO-SIC in \pyflosic{}~\cite{Schwalbe2020_JCP_84104, Liebing2022__167}, which is built on top of the \pyscf{}~\cite{Sun2020_JCP_24109} electronic structure package; the implementation of E-FLO-SIC used in this work is freely available on \gitlab{}.\bibnote{\pyflosic{} containing the employed E-FLO-SIC implementation is available at \url{https://gitlab.com/opensic/pyflosic2} (accessed 2024-04-17).}
Exemplifying the power of reusable software,\cite{Lehtola2023_JCP_180901} our E-FLO-SIC implementation in \pyflosic{} inherits all the features that \pyscf{} offers; note that as the required interfaces to the underlying quantum chemistry program are limited, the implementation could also be interfaced with other programs featuring a \python{} interface, such as the \psifour{}\cite{Smith2020_JCP_184108} or \chillipy{}\cite{Schwalbe2023_Z} programs.

Perhaps the most important feature of \pyscf{} for the present purposes is that any exchange-correlation functional in the \libxc{}~\cite{Lehtola2018_S_1} library that belongs to the local density approximation (LDA), generalized-gradient approximation (GGA), or meta-GGA level of Jacob's ladder\cite{Perdew2001_ACP_1} can be used in \pyflosic{} through \pyscf{}.
Despite the flexible support for various DFAs, in analogy to our previous work in \citeref{Trepte2021_JCP_224109}, all calculations discussed in the main text are carried out with the LDA of Perdew and Wang (SPW92),\cite{Bloch1929_ZfuP_545, Dirac1930_MPCPS_376,  Perdew1992_PRB_13244} as this level of theory is already sufficient to demonstrate symmetry breaking in FLO-SIC and the lack thereof in E-FLO-SIC.

For completeness, further calculations were performed utilizing the Perdew--Burke--Ernzerhof (PBE) GGA\cite{Perdew1996_PRL_3865, Perdew1997_PRL_1396} and the Tao--Perdew--Staroverov--Scuseria (TPSS) meta-GGA\cite{Tao2003_PRL_146401, Perdew2004_JCP_6898} functionals
to demonstrate the applicability of the proposed E-FLO-SIC method to higher levels of theory.
As the results of these calculations are analogous to those obtained with the SPW92 LDA functional, the PBE and TPSS data are not discussed in the main text but are available in the Supporting Information.

\pyscf{} also offers a variety of Gaussian-type orbital (GTO) basis sets, supporting basis functions up to angular momentum $l=15$ as well as effective core potentials for heavy elements;\cite{Sun2024_JCP_174116} these functionalities are all available in \pyflosic{}, as well.
As the symmetry breaking effects can be reproduced even in a minimal basis set, the double-$\zeta$ polarization consistent pc-1 basis set~\cite{Jensen2001_JCP_9113} was employed for all calculations for simplicity.
The SCF convergence threshold was set to $10^{-8}\,E_{\text{h}}$ for the total energy, unless specified otherwise.
An unpruned \pyscf{} grid level of 7 was used for the exchange-correlation quadrature with the multi-center scheme of \citeit{Becke1988_JCP_2547}, corresponding to a (90,974) grid for hydrogen and a (135,1202) grid for carbon following the approach of \citeit{Treutler1995_JCP_346}.

We implemented E-FLO-SIC in a two-step SCF cycle~\cite{Karanovich2021_JCP_14106}:  in each inner step, the FODs for each configuration are (re)optimized for a fixed electron density, which is then updated in the outer loop based on the SIC computed with fixed values of the FODs.
FOD configurations were initialized for each molecular geometry with the Monte Carlo-based method implemented in \fodmc{}.\cite{Schwalbe2019_JCC_2843}
Unless specified otherwise, the FODs were self-consistently optimized in (E-)FLO-SIC calculations such that the final maximum force $F_{\text{max}}$ on the FODs was below $1.5\times10^{-4}\,E_{\text{h}}/a_{0}$.
All calculations in this work were carried out within the spin unrestricted formalism. 

\subsection{Electronic Geometries \label{sec:elgeom}}

\begin{figure*}
	\centering
    \begin{subfigure}{\textwidth}
		\centering
		\includegraphics[width=0.24\textwidth]{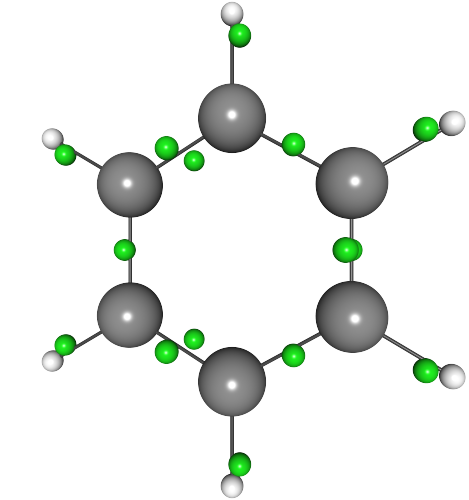}
		\includegraphics[width=0.24\textwidth]{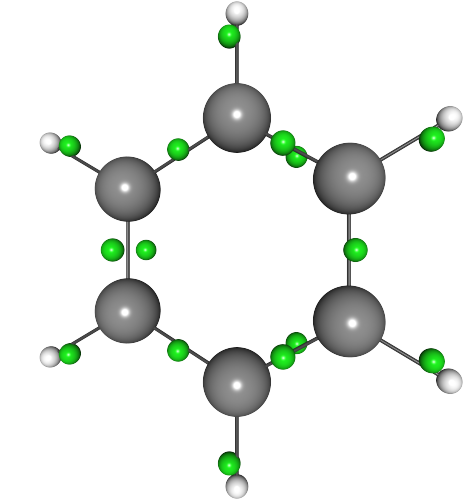}
		\includegraphics[width=0.24\textwidth]{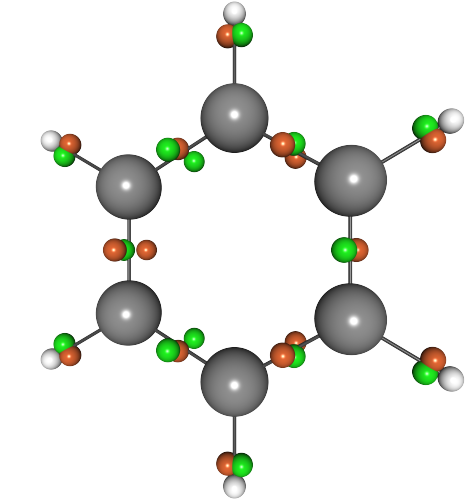}
		\includegraphics[width=0.24\textwidth]{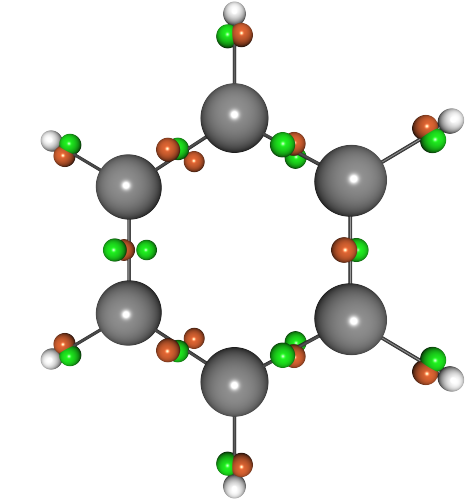}
		\subfloat[\label{fig:lt1} LT1\phantom{QQQ}]{\hspace{0.24\textwidth}}
		\subfloat[\label{fig:lt2} LT2\phantom{QQQ}]{\hspace{0.24\textwidth}}
		\subfloat[\label{fig:ldq1} LDQ1]{\hspace{0.24\textwidth}}
		\subfloat[\label{fig:ldq2} LDQ2]{\hspace{0.24\textwidth}}
	\end{subfigure}
    \caption{Optimized FOD configurations of benzene using the SPW92 LDA functional, visualized with \pyflosicgui{}~\cite{Schwalbe2020_JCP_84104,Liebing2022__167}.
        Carbon atoms are colored in grey, hydrogens in white, and the FODs in green and red denoting spin-up and spin-down FODs.
        In case the FOD positions in $\alpha$ and $\beta$ spin channels are identical, only the green FODs are visualized.}
    \label{fig:C6H6_configurations}
\end{figure*}

It has been previously shown that Linnett double-quartet (LDQ)~\cite{Linnett1960_N_859, Linnett1961_JACS_2643, Luder1966_JCE_55} theory yields suitable FOD configurations for describing aromatic systems like benzene (\ce{C6H6}) in FLO-SIC---with the drawback of spin symmetry breaking.\cite{Trepte2021_JCP_224109}
As we have discussed in \citeref{Trepte2021_JCP_224109}, one can think of four FOD configurations guided by chemical bonding theory for the benzene molecule: two from Lewis' theory (LT) of bonding~\cite{Lewis1916_JACS_762}, and another two from LDQ theory.
The two LT structures arise from the two ways to choose the single and double bonds in benzene: as illustrated by LT1 and LT2 in \cref{fig:C6H6_configurations}, LT predicts alternating \ce{C\bond{-}C} single and \ce{C\bond{=}C} double bonds with no spin polarization.
In contrast, LDQ theory employs staggered LT structures for the spin-up and spin-down channels $\alpha$ and $\beta$: LDQ1 = LT1$\alpha$ + LT2$\beta$ and LDQ2 = LT2$\alpha$ + LT1$\beta$ in \cref{fig:C6H6_configurations}.
LDQ thus achieves a constant bond order of 1.5 at the cost of breaking spin symmetry.

In addition to FLO-SIC calculations with the LT and LDQ structures, we also studied calculations based on ensembles of the FOD configurations, which we denote as \{LT1,LT2\} and \{LDQ1,LDQ2\}.
The use of the LT or LDQ configurations to form an ensemble for benzene is a special case of the general rule of combining all possible Lewis resonance structures for the molecule in analogy to classical chemical bonding theories; the generalization to other systems is therefore obvious.

Importantly, the functional defined by \cref{eq:ensemble-si,eq:EFLOSIC-tot} reduces to averaging the correction over the FOD structures of the molecule for each spin channel.
When a symmetric FOD structure (like the guesses produced by \fodmc{}) is employed, the LDQ structures are obtained by employing alternate sets of LT FODs for the two spins.
In this case, the \{LT1,LT2\} and \{LDQ1,LDQ2\} ensembles are mathematically equivalent, as \cref{eq:ensemble-si,eq:EFLOSIC-tot} will yield exactly the same ensemble solution for either case.

The FODs employed in the \{LT1,LT2\} and \{LDQ1,LDQ2\} ensembles can also be different, for example when the ensemble optimization is started from pre-optimized LT and LDQ geometries.
However, as we demonstrate in \cref{sec:initguess}, even in this case the two ensembles converge to the same final energy.

\subsection{Molecular Geometries}

All CH bond lengths were fixed at the experimental value $d_{\text{CH}} = 1.084$\,\AA{} \cite{Johnson2002}.
We considered symmetric molecular geometries with three CC bond lengths $d_{\text{CC}}$, which have been found to be minima in DFT and FLO-SIC calculations based on various FOD configurations, respectively~\cite{Trepte2021_JCP_224109}: $d_{\text{CC}} = 1.392$\,\AA{}, $d_{\text{CC}} = 1.377$\,\AA{}, and $d_{\text{CC}} = 1.362$\,\AA{}.

We also considered distorted geometries that feature distinct \ce{C\bond{-}C} and \ce{C\bond{=}C} bond lengths instead of a single CC bond length as in the symmetric geometries, in analogy to the work of \citeit{Trepte2021_JCP_224109}.
Seven local distortions $\Delta$ between 0.01\,\AA{} and 0.07\,\AA{} were investigated in steps of 0.01\,\AA{} for each of the three starting symmetric structures, $\Delta=0$ being also implicitly included by the symmetric structure.
The local distortions were chosen such that they elongate the three \ce{C\bond{-}C} bonds and shorten the three \ce{C\bond{=}C} bonds in LT1; the H atoms were moved accordingly.
Based on our knowledge from previous studies,\cite{Lehtola2016_JCTC_3195, Trepte2021_JCP_224109} this choice to study only the non-negative half-axis for the local distortion $\Delta$, $\Delta \ge 0$, is sufficient, since the results for $\Delta \to -\Delta$ are obtained by interchanging LT1 with LT2, and LDQ1 with LDQ2.

\section{Results \label{sec:results}}

As discussed in \cref{sec:compdec}, we will only discuss results obtained with the SPW92 functional in the main text.
Fully analogous results were also obtained with the PBE and TPSS functionals, as can be seen in the corresponding figures and tables included in the Supporting Information.

\subsection{Symmetry Breaking of Mo\-lec\-u\-lar Geometry \label{sec:molgeom}}

\begin{figure*}
	\centering
    \includegraphics[width=\textwidth]{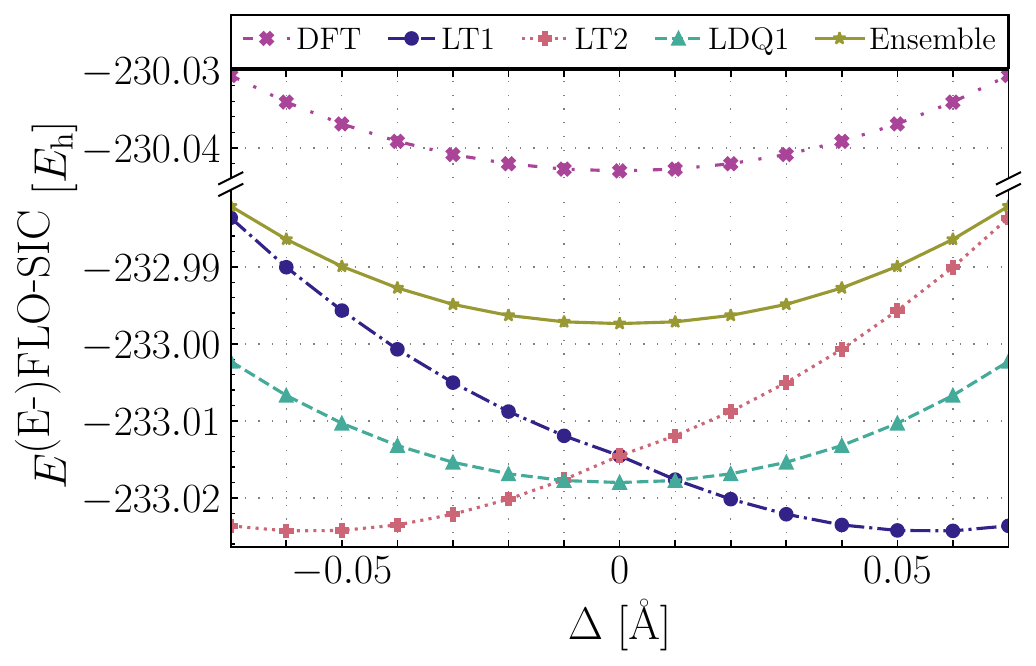}
    \caption{FLO-SIC and E-FLO-SIC total energies as a function of the local distortion $\Delta$ using the SPW92 LDA functional, visualized with \matplotlib{}~\cite{Hunter2007}.
      Data for the LDQ2 calculation is not shown, because it is indistinguishable from the shown LDQ1 data.
      As the \{LT1,LT2\} and \{LDQ1,LDQ2\} calculations were initialized from equivalent \fodmc{} starting points (see discussion in \cref{sec:elgeom}), their results are indistinguishable and have been marked here as ``Ensemble''.
    In-depth details on the minima of each calculation are given in \cref{tab:C6H6_configurations}.
    }
    \label{fig:C6H6_pes}
\end{figure*}

\begin{table*}
    \centering
    \caption{Properties of the optimum geometry of benzene and the corresponding wave function in FLO-SIC and E-FLO-SIC calculations with \pyflosic{} using the SPW92 LDA functional.
    The optimal bond lengths of short and long bonds $d^\text{short}_\text{CC}$ and $d^\text{long}_\text{CC}$ coincide in the case of a symmetric optimum geometry, but differ for a symmetry-broken optimal geometry.
    The value of the KS total energy $E^\text{KS}$ is minimized in the DFT calculation, but competes with the SIC in (E-)FLO-SIC calculations that minimize the total energy $E^\text{(E-)FLO-SIC}$, instead.
    For reference, the $\braket{\hat{\bm{S}}^2}$ value is also shown.}
    \label{tab:C6H6_configurations}
    \begin{tabular}{cccccc}
        \toprule
        & $d^\text{short}_\text{CC}$ [\AA{}] & $d^\text{long}_\text{CC}$ [\AA{}] & $E^\text{KS}$ [$E_\text{h}$]  & $E^{\text{(E-)FLO-SIC}}$ [$E_\text{h}$] & $\braket{\hat{\bm{S}}^2}$ \\
        \midrule
        DFT           & $1.392$ & $1.392$ & $-230.042950$ & - & $0.000$ \\
        \midrule
        \multicolumn{6}{c}{Configuration/Ensemble} \\
        \midrule
        LT1/LT2       & $1.317$ & $1.407$ & $-229.993022$ & $-233.024284$ & $0.000$ \\
        LDQ1/LDQ2     & $1.362$ & $1.362$ & $-229.998131$ & $-233.018003$ & $0.149$ \\
        Ensemble      & $1.362$ & $1.362$ & $-230.015048$ & $-232.997362$ & $0.000$ \\
        \bottomrule
    \end{tabular}
\end{table*}

We begin with the discussion of the symmetry breaking of the molecular geometry.
Our main result is shown in \cref{fig:C6H6_pes} and \cref{tab:C6H6_configurations}.
FLO-SIC calculations with LT1 and LT2 type FODs result in symmetry breaking,\cite{Trepte2021_JCP_224109} which is evident by the non-zero values of the symmetry breaking distortion $\Delta$ corresponding to the distinct minima for LT1 and LT2 in \cref{fig:C6H6_pes}.
In contrast, the E-FLO-SIC ensemble solutions prefer the symmetric molecular geometry, and thus do not exhibit symmetry breaking.
Although FLO-SIC calculations with LDQ1 and LDQ2 type FODs also predict a symmetric optimal geometry,\cite{Trepte2021_JCP_224109} these calculations lead to spin symmetry breaking,\cite{Trepte2021_JCP_224109} which is obvious from the non-zero value of $\langle \hat{\mathbf{S}}^2 \rangle$ in \cref{tab:C6H6_configurations}, while E-FLO-SIC exhibits no spin symmetry breaking.

Even if the visibly higher value of the total energy in \cref{fig:C6H6_pes} for the ensemble calculations over the LT or LDQ solutions may appear problematic, we note that such an effect is completely expected from the restoration of symmetries in a situation where the underlying theory prefers to break the symmetry: imposing the restriction of maintaining the correct molecular symmetry---if only by averaging over all possible electronic geometries so that any net bias zeros out---still constitutes a penalty which by necessity results in a higher value of the total energy.

Remarkably, examination of the numerical values of the energies in \cref{tab:C6H6_configurations} demonstrates that the energy gain from symmetry breaking is negligible compared to the overall effect of the self-interaction correction for the studied SPW92 calculations: the E-FLO-SIC solutions exhibit a total energy that is almost $3\,E_\text{h}$ lower than that of the KS-DFT calculation, while the differences observed in the total energies of E-FLO-SIC and FLO-SIC calculations are orders of magnitude smaller.
Because the overall magnitude of the SIC is well known to be much more considerable for LDA than for GGA and meta-GGA functionals, which often even exhibit negative self-interaction energies for valence orbitals while for LDA orbital self-interaction energies tend to be always positive, such a remarkable separation of scales does not occur for the PBE and TPSS functionals, as can be observed from the data in the Supporting Information.

We note that FLO-SIC calculations yield the same energy for LT1 and LT2 calculations at the initial guess FODs at any symmetric geometry.
Also the LDQ1 and LDQ2 structures as well as the corresponding \{LT1,LT2\} and \{LDQ1,LDQ2\} ensemble calculations show the same guess energy.
With the bond length $d^\text{short}_\text{CC}=d^\text{long}_\text{CC}=1.362$\,\AA{}, all of these calculations yield the initial guess energy $-232.910121\,E_{\text{h}}$ with the employed \fodmc{} initial FODs.
The total energies for the calculations become dissimilar when the FODs are relaxed, the resulting total energies having been given in \cref{tab:C6H6_configurations}.

\subsection{Symmetry Breaking of Electron Density \label{sec:eldens}}

\begin{figure*}
    \begin{subfigure}{\textwidth}
		\centering
		\includegraphics[width=0.19\textwidth]{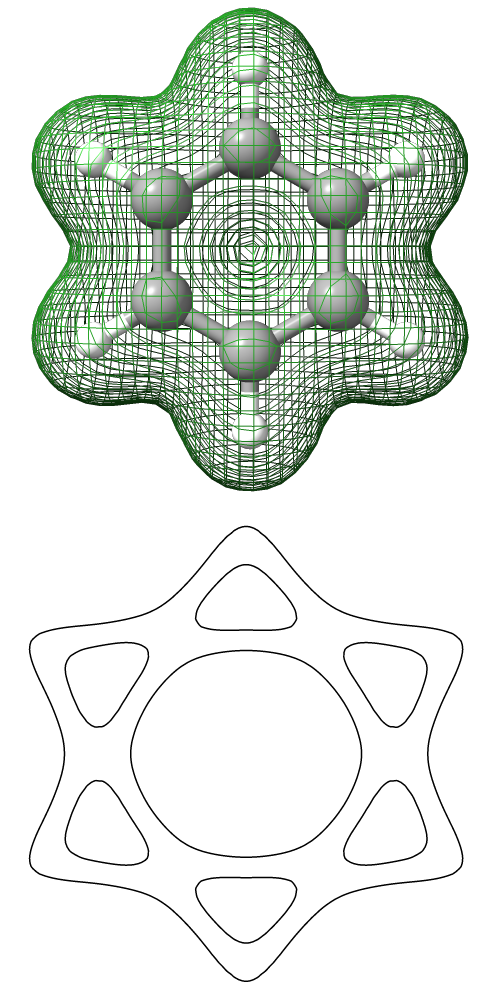}
		\includegraphics[width=0.19\textwidth]{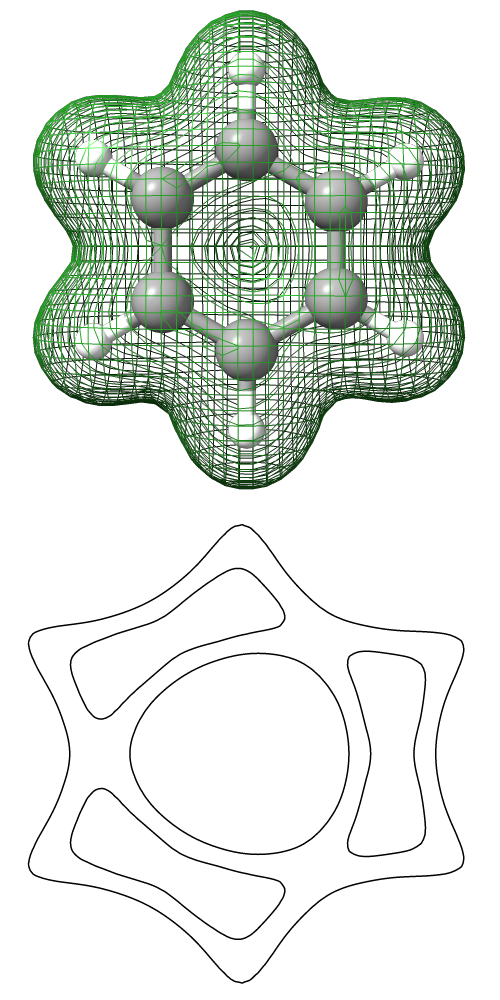}
		\includegraphics[width=0.19\textwidth]{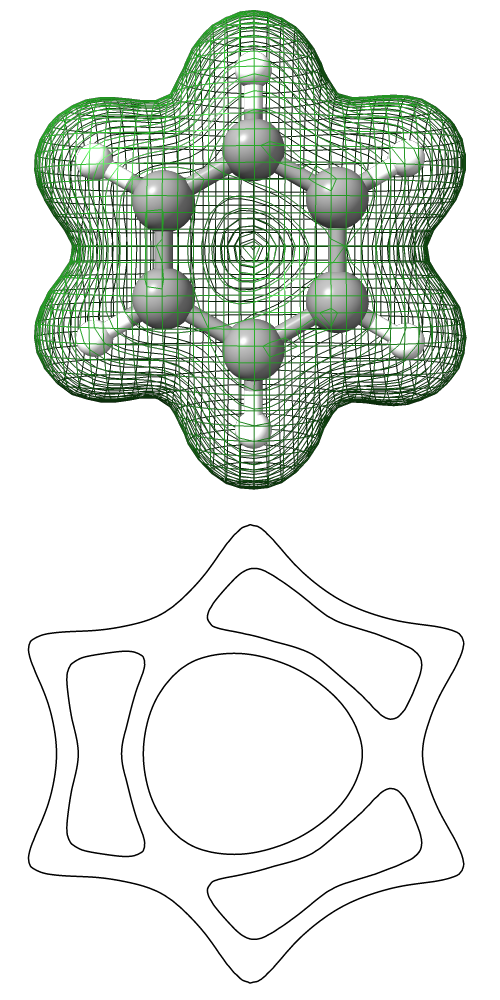}
		\includegraphics[width=0.19\textwidth]{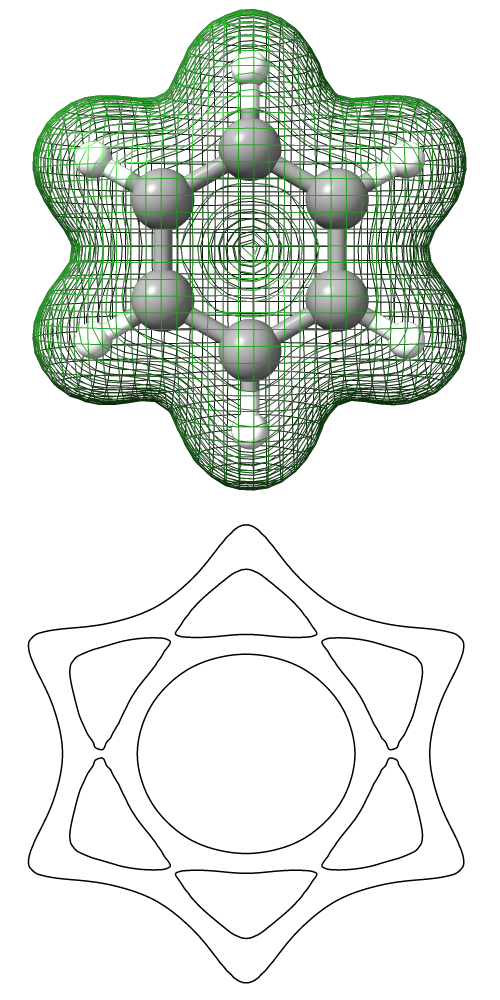}
        \includegraphics[width=0.19\textwidth]{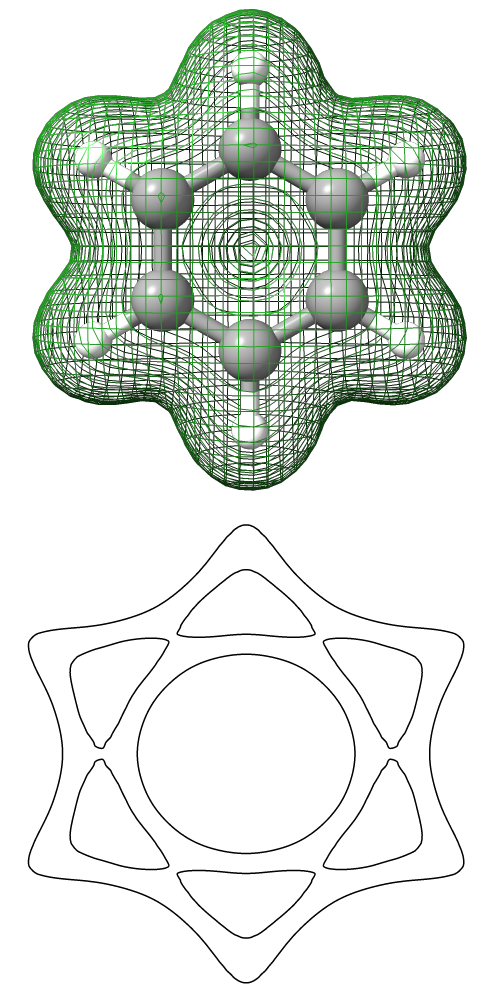}
		\subfloat[\label{fig:dft_den} DFT\phantom{QQQQ}]{\hspace{0.19\textwidth}}
		\subfloat[\label{fig:lt1_den} LT1\phantom{QQQ}]{\hspace{0.19\textwidth}}
        \subfloat[\label{fig:lt2_den} LT2\phantom{QQ}]{\hspace{0.19\textwidth}}
		\subfloat[\label{fig:ldq1_den} LDQ1\phantom{Q}]{\hspace{0.19\textwidth}}
		\subfloat[\label{fig:cc1_den} Ensemble]{\hspace{0.19\textwidth}}
    \end{subfigure}
    \caption{Wireframe plots (top) and contour plots (bottom) of the electron density of benzene from calculations using the SPW92 LDA functional, created with \chimerax{}~\cite{Pettersen2020}, and \plotly{}~\cite{Inc2015} and \eminus{}~\cite{Schulze2023a}, respectively. The geometries coincide with the ones found in \cref{tab:C6H6_configurations}. The isovalues of the wireframe plots have been chosen such that 95\,\% of the density is contained within the isovalue per the algorithm proposed by \citeit{Lehtola2014_JCTC_642}. The contour plots show the density 0.424\,\AA{} above the molecular plane. Only the density for LDQ1 is displayed since it is indistinguishable from the LDQ2 density.}
    \label{fig:C6H6_density}
\end{figure*}

We found in \citeref{Trepte2021_JCP_224109} that symmetry breaking may also be observed in the electron density, and demonstrated the effect on benzene with RSIC calculations initialized with FLOs corresponding to Lewis and Linnett's theories.
The relaxed FLO-SIC electron density breaks symmetry also at the symmetric molecular geometry, leading to the large artifactual dipole moments discussed in \citeref{Trepte2021_JCP_224109}.

We now demonstrate that the proposed ensemble method of this work reproduces a symmetric electron density.
The electron densities of benzene from the studied KS-DFT, FLO-SIC, and E-FLO-SIC calculations are shown in \cref{fig:C6H6_density}.
While clear symmetry breaking is observed in the contour plots of the electron density of the LT1 and LT2 calculations, the E-FLO-SIC calculations clearly reproduce an electron density that has the same symmetry as the molecule itself, like the KS-DFT electron density.
We note again that even though FLO-SIC calculations based on LDQ type FODs also result in a total electron density that has the same symmetry as the molecule, these calculations result in an artifactual spin density,\cite{Trepte2021_JCP_224109} which results in the non-zero value of $\langle \hat{\mathbf{S}}^2 \rangle$ discussed above and shown in \cref{tab:C6H6_configurations}.

\subsection{Dependence on Initial Guess \label{sec:initguess}}
As a final point, we show that the result of the ensemble calculation is not sensitive to the starting guess by repeating the calculations for the symmetric geometry, i.e., bond lengths $d^\text{short}_\text{CC}=d^\text{long}_\text{CC}=1.362$\,\AA{} from different FOD starting points.
While the calculations discussed above in \cref{sec:molgeom} and \cref{tab:C6H6_configurations} were begun from \fodmc{} FODs, for which the \{LT1,LT2\} and \{LDQ1,LDQ2\} ensembles coincide, now we investigate calculations for the two ensembles initialized with pre-optimized FODs for the LT1, LT2, LDQ1, and LDQ2 configurations.

For these calculations, we made the employed FOD optimization criterion $5$ times smaller than in \cref{sec:compdec}: we now demand that the final maximum force $F_{\text{max}}$ acting on the FODs is below $3\times10^{-5}\,E_{\text{h}}/a_{0}$.
We also ensured a tight convergence of the gradient norm of the orbitals (\pyscf{} parameter \verb|conv_tol_grad|=$1\times10^{-7}$), even though changing the SCF convergence criterion did not appear to affect the results.

Since the pre-optimized LT1, LT2, LDQ1, and LDQ2 FODs are distinct, the \{LT1,LT2\} and \{LDQ1,LDQ2\} ensemble calculations start off from slightly different total energies: $-232.910541\,E_{\text{h}}$ and $-232.910520\,E_{\text{h}}$, respectively.
These values are slightly lower than the initial energy discussed in \cref{sec:molgeom}, decreasing the PZ energy by $0.4\,\text{m}E_h$ compared to that obtained with the \fodmc{} initial guess.

Running the optimization, a significant decrease is observed in the values of the E-FLO-SIC energies of the two calculations, and an excellent agreement is reached: both calculations converge to the total energy of  $-232.997466\,E_{\text{h}}$.
An investigation of the corresponding log files reveals a difference of a mere $0.2\,\mu E_{\text{h}}$ between the two calculations, which is much smaller in magnitude than the employed FOD convergence threshold. 
We take this as proof that the optimizations have converged to the same minimum.

The attentive reader will by now have noted that these total energies are slightly different from the values discussed above in \cref{sec:molgeom}: a difference of around $0.1\,\text{m}E_{\text{h}}$ is observed from the values given in \cref{tab:C6H6_configurations}.
However, rerunning the calculation of \cref{tab:C6H6_configurations} with the convergence thresholds employed in this subsection leads to a total E-FLO-SIC energy of $-232.997467\,E_{\text{h}}$.
Without intermediate rounding, the total energies from different calculations agree to a precision of less than a microhartree.
Thus, E-FLO-SIC calculations do appear to converge to the same ensemble solution even when starting from different FODs.

\section{Summary and Discussion \label{sec:summary}}

We have solved a long-standing issue in the  Perdew--Zunger self-interaction correction\cite{Perdew1981_PRB_5048} (PZ-SIC) method: symmetry breaking caused by distinct local minima that correspond to distinct Kekul{\'e} structures for the system.
Our ensemble PZ-SIC (E-PZ-SIC) method is obtained by averaging the self-interaction correction over sets of localized orbitals that correspond to the various Kekul{\'e} structures / local minima.
The use of an ensemble of localized orbitals aligns perfectly with the chemical interpretation of resonance structures: a complete picture of chemistry is only obtained by including all possible structures in the calculation.
The optimizations of the various electronic configurations included in the ensemble are embarrassingly parallel, and the overall cost of the method grows only linearly in the number of electronic geometries.
As long as the number of necessary configurations does not explode, the method maintains the asymptotic scaling of the PZ-SIC method with system size.

We pointed out that E-PZ-SIC is straightforward to realize within the PZ-SIC approach based on Fermi--L{\"o}wdin orbitals (FLOs), FLO-SIC,\cite{Trepte2021_JCP_224109} as only minimal changes are needed to generalize an existing FLO-SIC implementation to the ensemble FLO-SIC (E-FLO-SIC) method.
We exemplified the E-FLO-SIC method with calculations with LDA, GGA, and meta-GGA functionals on the benzene molecule, which is well-known to result in symmetry breaking in PZ-SIC\cite{Lehtola2016_JCTC_3195} and FLO-SIC.\cite{Trepte2021_JCP_224109}
We showed that E-FLO-SIC correctly produces symmetric electron densities and molecular structures with no spin polarization, while FLO-SIC predicts either symmetry-broken electron densities and molecular geometries or spin symmetry breaking, as we already showed in \citeref{Trepte2021_JCP_224109}.
We also showed that the E-FLO-SIC method converges to the same minimum whether the employed ensemble started off symmetric or not.
We are therefore confident that we have solved the issues plaguing  PZ-SIC and FLO-SIC calculations of aromatic systems, for which PZ-SIC and FLO-SIC solutions often exhibit large artifactual dipole moments, for instance.\cite{Trepte2021_JCP_224109}

Further research is planned to investigate if E-PZ-SIC or E-FLO-SIC can consistently cure or at least reduce further problems of SIC, for instance, the overestimation of ionization potentials, and unsatisfactory predictions of enthalpies of formation.
We hope to follow up this proof-of-concept study with more in-depth applications of E-PZ-SIC and E-FLO-SIC.

Finally, it is known that some of the shortcomings of PZ-SIC and FLO-SIC can be improved by scaling down the correction in many-electron regions.\cite{Vydrov2006_JCP_094108, Vydrov2006_JCP_191101, Zope2019_JCP_214108, Bhattarai2020_JCP_214109} 
Because the scaled-down methods maintain a mathematical similarity to PZ-SIC and FLO-SIC, likewise relying on optimization of localized orbitals to evaluate the SIC, we expect them to have similar issues with multiple local minima, as well. 
However, continuing the analogy, we also expect the E-PZ-SIC / E-FLO-SIC method to generalize to those types of methods without any issues.

By the request of a reviewer, we end the discussion with a brief summary of studies related to symmetry breaking in DFT.
Spin symmetry breaking of the spatial orbitals is typically used in the literature to obtain more reliable energy estimates for chemical reactions that involve breaking chemical bonds, as spin-unrestricted calculations allow many types of bonds to dissociate correctly.
However, spin symmetry breaking has also been observed to help in certain cases near the equilibrium bond length.
For example, \citeit{Perdew2023_JPCA_384} recently found that breaking the spin symmetry of the spatial orbitals leads to improved estimates for the atomization energy of \ce{C2} with the SCAN functional,\cite{Sun2015_PRL_36402} since unrestricting the spin is found to lead to a lower energy at the optimal internuclear distance.
However, the same procedure was found to lead to poorer agreement for the PW92 and PBE functionals that overestimate the atomization energy already when the spin symmetry is not broken.
The SCAN functional also predicts the \ce{Cr2} molecule to be spin polarized, and yields a potential energy curve which is in poorer agreement with experiment than the PBE curve.\cite{Zhang2020_PCCP_19585, Zhang2020_PCCP_24813}

Extremely recent work by \citeit{Maniar2024_JCP_144301} suggests that the \emph{net} spin on the two Cr atoms in \ce{Cr2} can be eliminated in PW92-FLO-SIC by a symmetry broken FOD structure, with oscillations in the spin polarization integrating to close to zero.
However, finding the true ground state configuration is often painstaking in FLO-SIC,\cite{Trepte2021_JCP_224109} and we expect transition metal systems to be especially challenging.
Indeed, many kinds of FOD geometries can be found for transition metal atoms like Cr,\cite{Kao2017_JCP_164107, Maniar2024_JCP_144301} and it is at the moment not clear how well self-interaction corrections work for such challenging systems. 

We also note that the SCAN functional also incorrectly predicts a symmetry broken ground state for graphene and benzene.\cite{Zhang2020_PCCP_19585, Zhang2020_PCCP_24813}
However, SCAN is also well-known for its poor numerical behavior,\cite{Bartok2019_JCP_161101, Furness2020_JPCL_8208, Lehtola2022_JCP_174114} and the previous studies did not report whether the regularized rSCAN\cite{Bartok2019_JCP_161101} and r$^2$SCAN\cite{Furness2020_JPCL_8208} functionals suffer from similar issues.
Furthermore, we note that the optimal DFA for PZ-SIC calculations may differ from the DFA that is optimal for plain KS-DFT,\cite{Jonsson2015_PCS_1858, Lehtola2016_JCTC_4296} and that improved results may be obtained by scaling down the SIC in many-electron regions globally \cite{Vydrov2006_JCP_094108, Vydrov2006_JCP_191101} or locally \cite{Zope2019_JCP_214108}.
We expect that the development of various SICs will continue in the near future with studies to address the aforementioned list of issues affecting various DFAs, and remind the reader once again that the ensemble approach proposed in this work can equally well be applied to any new types of DFAs as well as globally or locally scaled-down versions of PZ-SIC.

We conclude by noting that symmetry breaking is also an issue in PZ-SIC calculations based on approximate localized orbitals, such as in the recent implementation of  \citeit{Peralta2024}.
However, our ensemble method can also be applied in such contexts, as well.

\citeit{Bi2024_JCP_34106} have recently proposed a method that unites variational optimization of the PZ-SIC functional with the von Niessen orbital localization method.\cite{Niessen1972_JCP_4290}
\citeit{Bi2024_JCP_34106} find that such a combination reduces the number of local solutions to the PZ-SIC equations.
Since various orbital localization methods are all known to yield distinct sets of localized orbital solutions for systems such as the benzene molecule studied in this work, we believe that our method will also be useful in the context of methods following \citeit{Bi2024_JCP_34106}.


\begin{acknowledgement}
This work contributes to our OpenSIC project which aims to bring
new perspectives to the field of self-interaction corrections
through the use of open-source software.
The authors thank the Universit{\"a}tsrechenzentrum of the Friedrich Schiller
University Jena for computational time and support.
S. S. was supported by the European Research Council (ERC) under the European Union’s Horizon 2022 research and innovation program (Grant agreement No. 101076233, ``PREXTREME'') and by the Center for Advanced Systems Understanding (CASUS)
which is financed by Germany’s Federal Ministry of Education and
Research (BMBF) and by the Saxon state government out of the State
budget approved by the Saxon State Parliament.
Views and opinions expressed are however those of the authors only and do not necessarily reflect those of the European Union or the European Research Council Executive Agency. Neither the European Union nor the granting authority can be held responsible for them.
Computations were performed at the Norddeutscher Verbund f{\"u}r Hoch- und H{\"o}chstleistungsrechnen (HLRN) under grant mvp00024.
W. T. S. highly acknowledges funding by the Deutsche Forschungsgemeinschaft (DFG,
German Research Foundation) - Research unit FuncHeal, project ID
455748945 - FOR 5301 (project P5). 
S. L. thanks the Academy of Finland for financial support under project numbers 350282 and 353749.
\end{acknowledgement}

\begin{suppinfo}
The results utilizing the PBE GGA functional and the TPSS meta-GGA can be found in the supporting information.
The employed molecular geometries and the optimized sets of FODs are available at \url{https://gitlab.com/opensic/ensemble_supplementary}.
\end{suppinfo}

\bibliography{merged.bib}  

\end{document}